\documentclass{SCIS2018}
 \usepackage{etex}
\usepackage{pgfplots}
\usepackage{tikz}
\usepackage{ifthen}
\usepackage{pifont}
\usepackage{epstopdf}
\begin{document}
    
    \ArticleType{RESEARCH PAPER}
    \Year{2018}
    \Month{}
    \Vol{61}
    \No{}
    \DOI{}
    \ArtNo{}
    \ReceiveDate{}
    \ReviseDate{}
    \AcceptDate{}
    \OnlineDate{}
    
    \title{Real-time intelligent big data processing: technology, platform, and applications}{Real-time intelligent big data processing: technology, platform, and applications}
    
    \author{Tongya ZHENG}{}
    \author{Gang CHEN}{}
    \author{Xinyu WANG}{}
    \author{Chun CHEN}{}
    \author{Xingen WANG}{{newroot@zju.edu.cn}}
    \author{Sihui LUO}{}
    
    \AuthorMark{Zheng T Y}
    
    \AuthorCitation{Zheng T Y, Chen G, Wang X Y, et al}
    
    
    \address[1]{College of Computer Science and Technology, Zhejiang University, Hangzhou {\rm 310027}, China}
    
    \abstract{Human beings keep exploring the physical space using information means. Only recently, with the rapid development of information technologies and the increasing accumulation of data, human beings can learn more about the unknown world with data-driven methods. Given data timeliness, there is a growing awareness of the importance of real-time data. There are two categories of technologies accounting for data processing: batching big data and streaming processing, which have not been integrated well. Thus, we propose an innovative incremental processing technology named after Stream Cube to process both big data and stream data. Also, we implement a real-time intelligent data processing system, which is based on real-time acquisition, real-time processing, real-time analysis, and real-time decision-making. The real-time intelligent data processing technology system is equipped with a batching big data platform, data analysis tools, and machine learning models. Based on our applications and analysis, the real-time intelligent data processing system is a crucial solution to the problems of the national society and economy.}
    
    \keywords{batching big data, streaming processing technology, real-time data processing, incremental computation, intelligent data processing system}
    
    \maketitle

    \section{Introduction}
    The cyberspace has emerged since people communicate with each other by information means. Decades ago, telegram, the telephone, and sensing devices were invented to assist human begins with communication and exploration. Recently, the cyberspace has been expanding greatly, because mobile terminals and the Internet could be easily accessed by billions of people. Thus, the world has evolved into a ternary CPH space(cyber, physics, and human society)~\cite{1}. 
    
    \begin{figure}
        \centering
        \begin{minipage}{.35\textwidth}
            \centering
            \includegraphics[width=\linewidth]{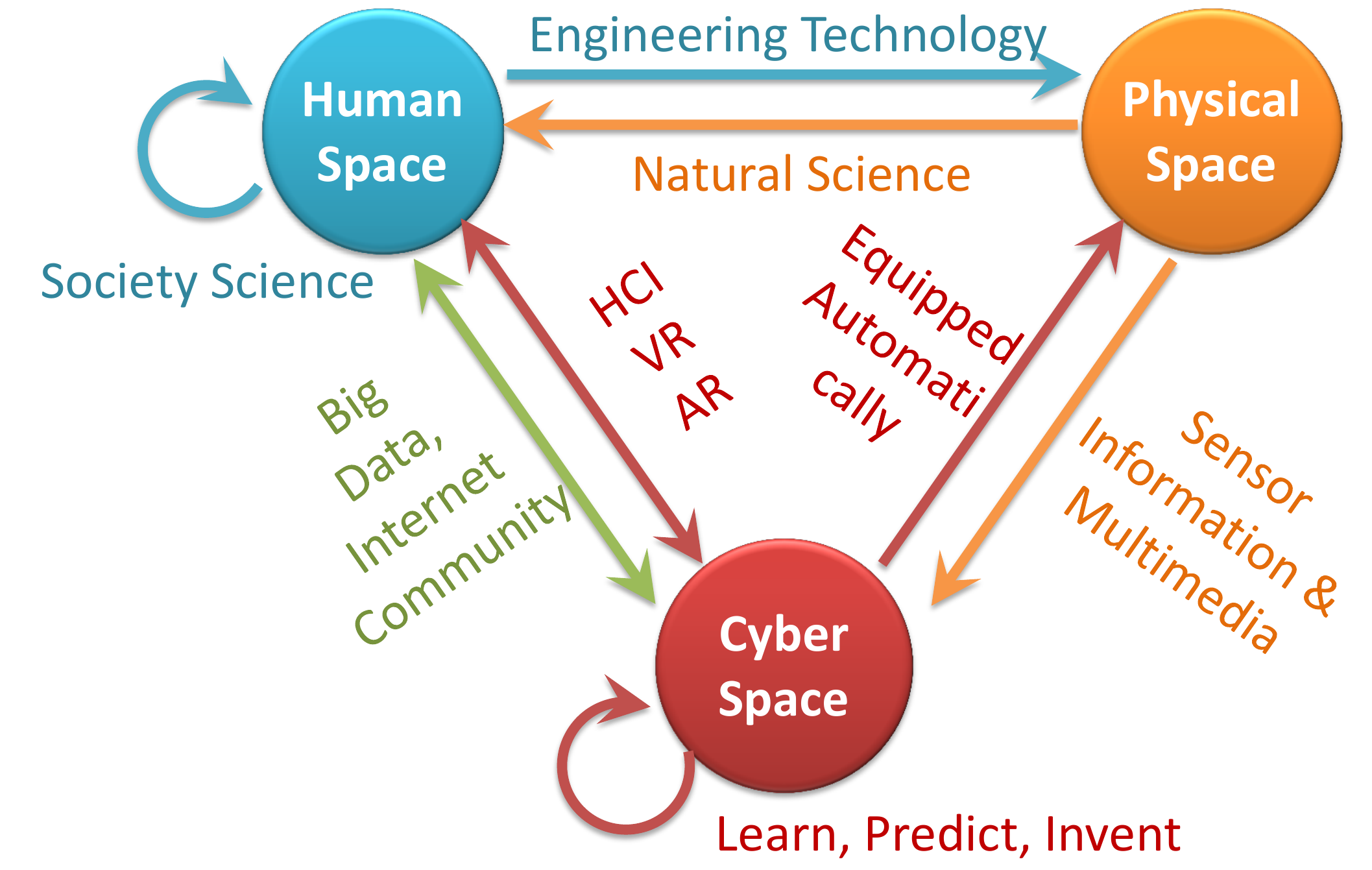}
            \captionof{figure}{CPH Space}
            \label{fig:cph}
        \end{minipage}%
        \begin{minipage}{.52\textwidth}
            \centering
            \includegraphics[width=\linewidth]{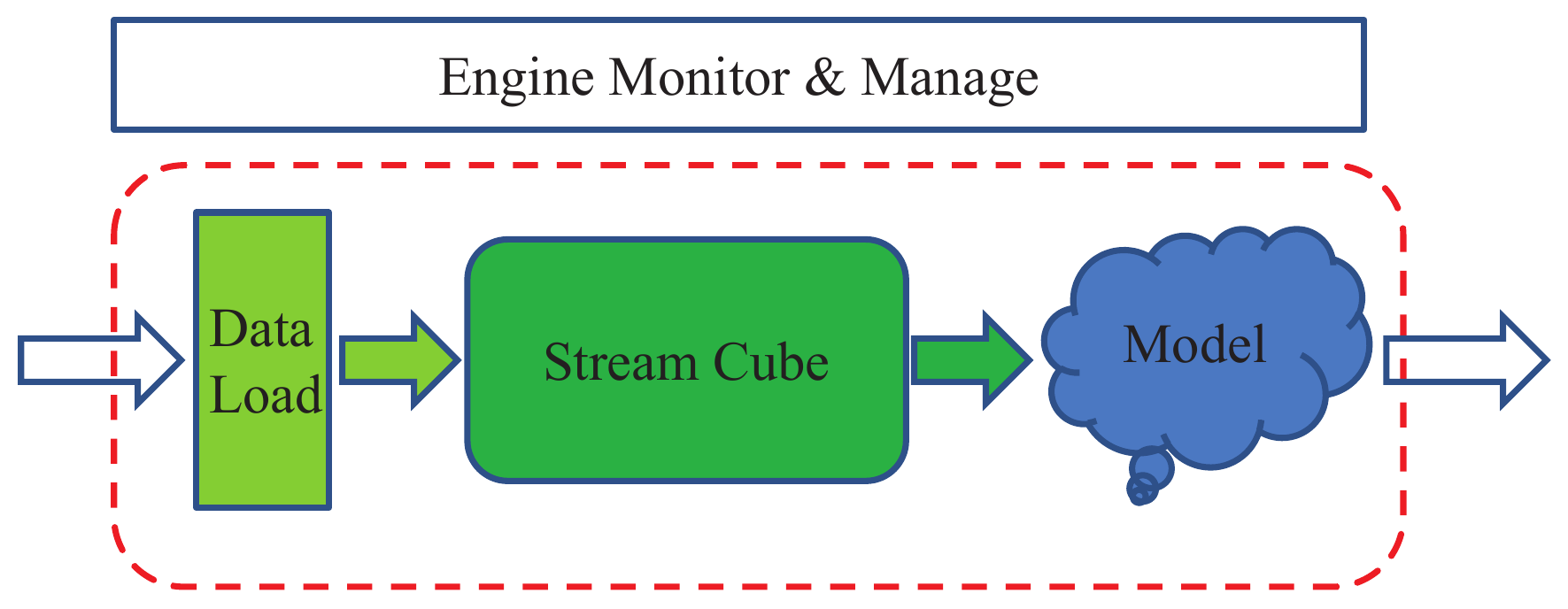}
            \captionof{figure}{The architecture of the intelligent system}
            \label{fig:system-arch}
        \end{minipage}
    \end{figure}
    
    As shown in Figure \ref{fig:cph},  the three space, the human space, the cyberspace, and the physical space, interact with each other more frequently today. Nowadays, the cyberspace has received massive sensor information and multimedia data, with numerous sensing devices in the physical space. Besides, on the one hand, human space exchanges information via the cyberspace. On the other hand, for the convenience of human beings, human-computer interaction and augmented reality have been developed.
    
    However, with the expansion of the cyberspace, petabytes of data emerge each day. Not only do we need to complete business logic in such a high concurrency scenario, but also we have to mine latent patterns of users to provide better service. For processing big data, there are two categories of systems: batching big data and streaming processing technology~\cite{2}. For mining latent preference of users' behavior, there is an increasing demand for applying artificial intelligence. In consideration of existing data processing platforms, it is hardly sufficient to realize real-time big data processing. What is more, there is a lack of artificial intelligence for most big data systems.
    
    Currently, data technologies have been broadly adopted because of the importance of the data. Although the problem of processing hundreds of gigabytes of data has been solved, it is recognized that data produced ten years ago is far less valuable than data in an hour. It means that the more recent data is, the more value it owns. Thus, the so-called real-time data matter most, as they are usually produced a few seconds from now, or even less than one millisecond. 
    
    To handle the real-time big data, we propose an innovative incremental data processing technology named after Stream Cube. To reach the goal, we utilize an incremental computation scheme and polynomial decomposition method, which decomposes complex statistical metrics into a sequence of mid-term representations. Thus, we could support statistical metrics computation over any time interval or real-time. 
    
    Furthermore, as shown in Figure \ref{fig:system-arch}, we implement a competitive real-time intelligent data processing system. The first key component of the system is the Stream Cube, accounting for real-time big data processing. The second key is the AI(Artificial Intelligence) Model module, responsible for mining patterns of the behaviour data of users. Lastly, before the Stream Cube, we provide the Data Load module for the combination of historical data and real-time data. Moreover, we provide the performance of the system via two practical applications. The one is the real-time risk detection in China UMS; the other is telling computers and humans apart in 12306 railway ticket website.
    
    To summarize, our main contributions are:
    \begin{itemize}
        \item We propose an innovative incremental real-time big data processing technology named after Stream Cube.
        \item  We also implement a real-time intelligent data processing system, which has not been well-studied yet.
        \item Finally, we analyze two real-time big data applications and evaluate the intelligent data processing system in these scenarios.    
    \end{itemize}
    
    The rest of the paper is structured as follows. In Section 2, we review the history of data processing and summarize recent advances in streaming processing. Section 3 firstly describes four challenges for real-time streaming processing and presents our innovative technology, the Stream Cube. Then, we implement a dual-core real-time intelligent system, whose dual-core are the Stream Cube and the AI Model module. The experimental results of the Stream Cube and analysis are presented in Section 4. We demonstrate two practical applications of the intelligent system in Section 5. Finally, Section 6 concludes our work. 
    
    \section{Big data processing: batching and streaming}
    
    \subsection{History of data processing}
    
    Date back to 1970s, the relational database was invented, which conforms to the principles of ACID(Atomicity, Consistency, Isolation, Durability). Besides, it could be managed with SQL(Standardized Query Language). Later in the 1990s, the data warehouse played an essential role as a system integrating data acquisition, data storage, and data access. However, since 2006, it has been replaced by big data processing technology, also called batching data processing technology. Compared to the small data processing capacity of the relational database in the 1970s, the data warehouse can handle far more data. However, practitioners had to endure extremely high latency for several days or tens of hours with the data warehouse. As for the batching data processing technology nowadays, it takes only several hours even tens of minutes to process terabytes of data, which is many times faster than that of the data warehouse. 
    
    Big data processing systems, considering their timeliness, have typically been divided into two groups, namely batching big data systems and streaming processing systems. Among them, batching big data is also known as historical big data. Meanwhile, streaming processing systems are designed for handling stream data, also called real-time data. 
    
    The widely used big data technologies have significantly been adopted to stimulate the economic growth of many industries, including the Internet industry as well as the traditional industries. Currently, the typical big data processing technical architecture includes Hadoop~\cite{3} and its derived systems. The Hadoop technical architecture, mainly supported by Yahoo! and Facebook, implements and optimizes the MapReduce~\cite{4} framework. Since it had been released in 2006, the Hadoop technical architecture has evolved into a vibrant ecosystem containing over sixty components. Furthermore, the ecosystem has significantly promoted the research on distributed computation~\cite{5}, cloud computation~\cite{6, 7}, and its related works~\cite{8}. 
    
    In conclusion, there are two classes of data processing systems, the batching big data processing system and the stream data processing system. The batching big data processing system, represented by Hadoop technical architecture and Spark, supports flexible analysis patterns and has the capability of dealing with billions of records. Nevertheless, it would take high latency to calculate the results of stream data, so that it is unreliable for the batching data system to process stream data. In contrast to the cluster computation and distributed computation of the batching big data processing system, the stream data system accelerates the computation speed with the help of in-memory computing. The stream data processing system, for example, Tez~\cite{9}, Spark Streaming~\cite{10}, Storm~\cite{11}, and Flink~\cite{12}, obtains low latency and handles real-time stream data. However, it is hard for the stream data system to change analysis patterns. Also, the application of the stream data system is restricted due to its small processing data volumes. 
    
    \subsection{Recent advances in real-time streaming processing}
    
    The evolution of streaming processing comes from raising awareness of the importance of data. According to its timeliness, data can be divided into three kinds. We define warm data as the data of the previous day. Therefore, we define the data before the previous day as the historical data, also named cold data. While the definition of hot data is not precise, traditionally, the data in one day could even be hot data. However, for an e-commerce platform like Taobao, it is crucial to distinguish normal behaviors from wicked ones even in several milliseconds. As we have depicted before, it means that real-time data is the most valuable. As a result, the focus of the big data community has gradually turned to the real-time data processing technologies because of its higher value. With their efforts, there have been several stream data processing components like Tez~\cite{9} and Spark Streaming~\cite{13}. 
    
    The meaning of the real-time data, also called the stream data, is that thousands of data sources are producing the time-ordered unbounded data endlessly. The time-ordered unbounded data is needed to be processed incrementally according to its occurrence time. Since 2012, the streaming processing technologies have obtained a comparative low latency of a few or tens of milliseconds, sometimes even smaller than one millisecond. However, even if in-memory computing and dynamic strategies are used to relieve the computation overhead, the data volume for streaming processing is still restricted.
    
    In the view of ~\cite{14}, existing streaming processing systems have three versions: Data Stream Management System(DSMS), Complex Event Processing(CEP) system, and streaming processing platform. The three versions differ in their goals. DSMS behaves similarly as traditional DBMS(Database Management System). CEP is more productive than DSMS in analyzing the structured data. There have been several mature solutions, e.g., Oracle Event Processing, Microsoft StreamInsight~\cite{15}, SQLStream s-Server~\cite{16}, StreamBase~\cite{17}, Esper, Cayuga~\cite{18}, and Apama~\cite{19}. Nevertheless, researchers and engineers show more interest in a more general streaming processing methodology. There have already been a few real-time streaming processing platforms, e.g., Apache Storm~\cite{11}, Apache Samza~\cite{20}, Yahoo! S4~\cite{21}, Spring XD, Spark Streaming~\cite{10}. These platforms have taken a few steps towards a fast, scalable, and robust streaming processing system. In this field, research institutions and top-tier enterprises work together to make the systems more advanced.  
    
    In a more detailed viewpoint, streaming processing systems usually can be divided into two classes. Systems in ~\cite{13, 22} focus on the problem of streaming processing based on big data processing solutions. On the contrary, in ~\cite{12, 11}, systems are designed for stream data originally. Furthermore, researchers have dived into the streaming processing field and proposed problems in different views. ~\cite{23, 24} focus on the issue of frequently variable workload. Meanwhile, ~\cite{25} aims to reduce the adaptive overhead in a more stable environment. Then, ~\cite{26, 27} turn to the optimization strategy development due to the ubiquity of the wide-area data sources. More interestingly, Scotty~\cite{28} shares a similar idea with the Stream Cube, targeting at splitting streams into non-overlapping slices and then aggregating them for different queries.
    
    In January 2017, a unified programming framework, Apache Beam Project, was released, supporting a mixed programming scheme of batching big data and stream data. For all, it just provides a unified programming interface, but not a fusion of batching big data and stream data. Therefore, it is of great need to develop a fast, efficient and controllable streaming processing system.
    
    \section{Real-time big data processing and intelligent data processing}
    
    As aforementioned, two significant problems are remaining unsolved in real-time data processing. The first is the processing problem of real-time big data. Hence, we propose an incremental computation scheme to solve the problem in real time. The second is that existing systems are lack of AI ability. Therefore, we design a real-time intelligent system to bring powerful AI tools into the data processing field.

      \begin{figure}[!t]
        \centering
        \includegraphics[width=0.55\textwidth]{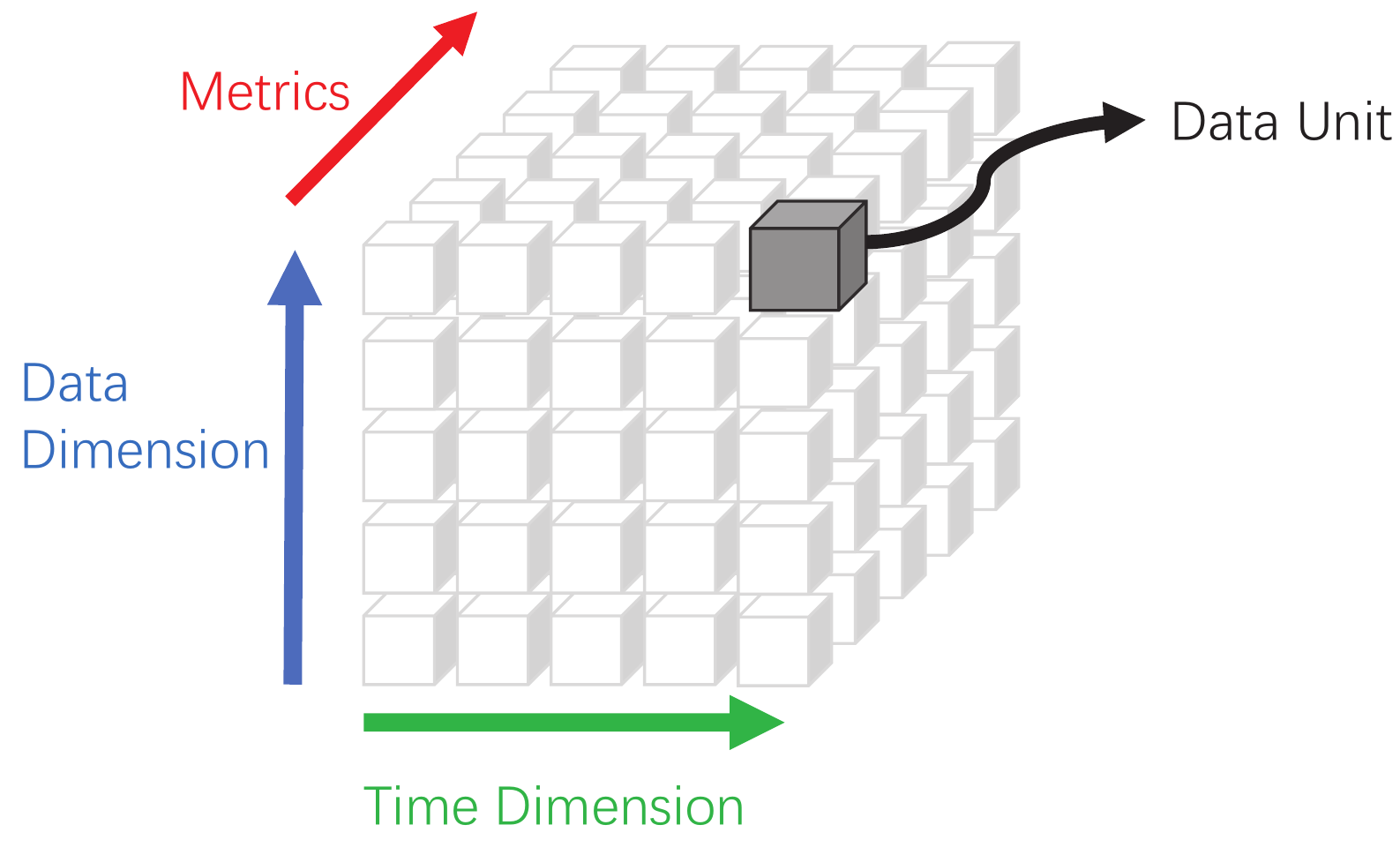}
        \caption{Computation dimensions of the Stream Cube}
        \label{fig:streamcube}
    \end{figure}
    
    \subsection{Real-time big data processing technology: Stream Cube}
    
    Since 2013, we have developed a stream big data processing platform, named the Stream Cube. As a productive platform, it focuses on the fusion of batching big data and stream data processing. Thus, it obtains the abilities of traditional batching processing and innovative streaming processing. Naturally, it supports queries of flexible time windows, driven by data and system clocks. What makes it useful for real-time data processing is that it utilizes a complex incremental algorithm to compute metrics defined in the time dimension and space dimension. The space dimension means that business experts define many metrics across primary statistical metrics of different dimensions. Moreover, as an extension, we connect the Stream Cube with statistical analysis and event sequence analysis.

    
    
\begin{figure}[!t]
        \centering
        \includegraphics[width=0.95\textwidth]{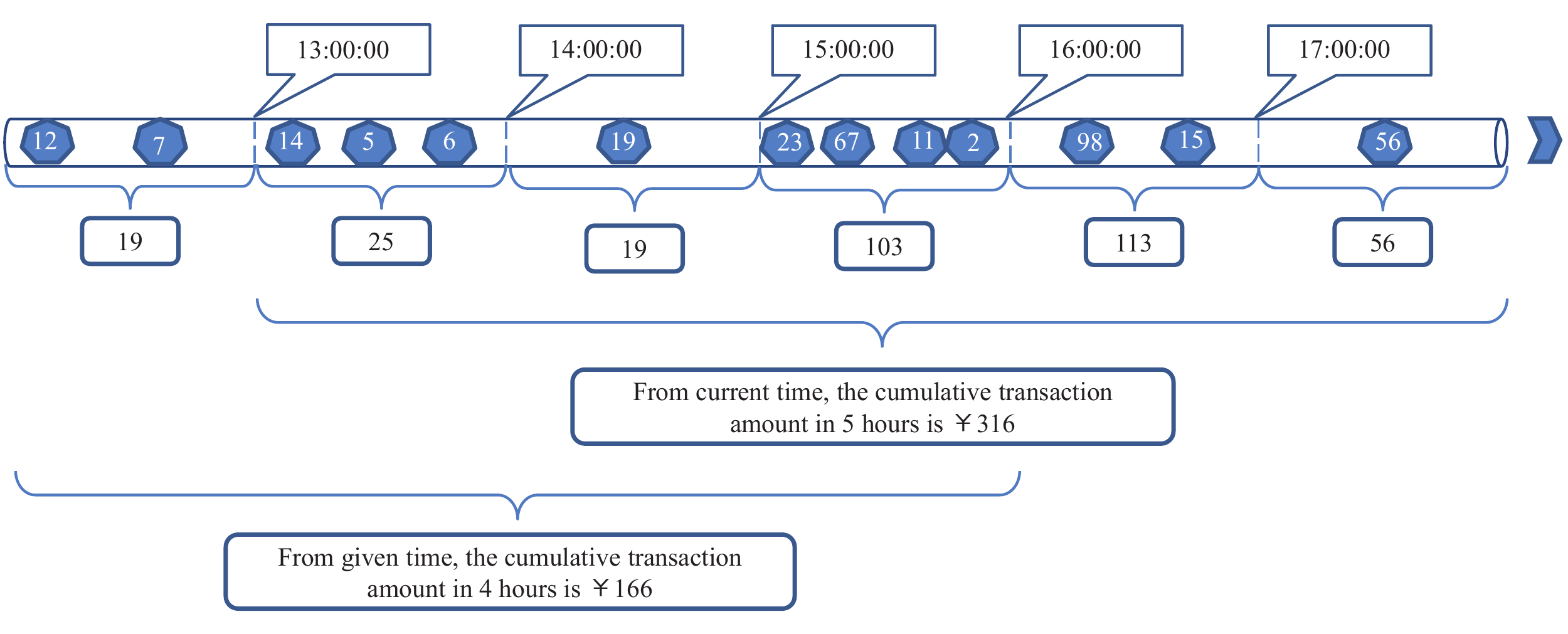}
        \caption{Challenge 2: different queries share an overlapping time interval}
        \label{fig:challenge2}
    \end{figure}

    
    In order to realize real-time big data processing, the Stream Cube has solved four challenges. The first challenge is the parallel computation based on distributed memory, which has been solved well by many frameworks like Aerospike~\cite{29} and Redis~\cite{30}. Secondly, there are two severe problems in the high-performance analysis of vast amounts of historical data. One is the redundant computation caused by time window shifts of different queries. Another is the difficulty of incremental calculation over vast amounts of historical data. For example, in Figure \ref{fig:challenge2}, the one query is the accumulative amount of transactions in 5 hours, and the other is the accumulative amount of transactions in 4 hours from 12 o'clock. In this example, the two queries share metrics of 3 hours that can be saved if the redundancy of computation can be eliminated. 
    
\begin{figure}[!t]
        \centering
        \includegraphics[width=0.6\textwidth]{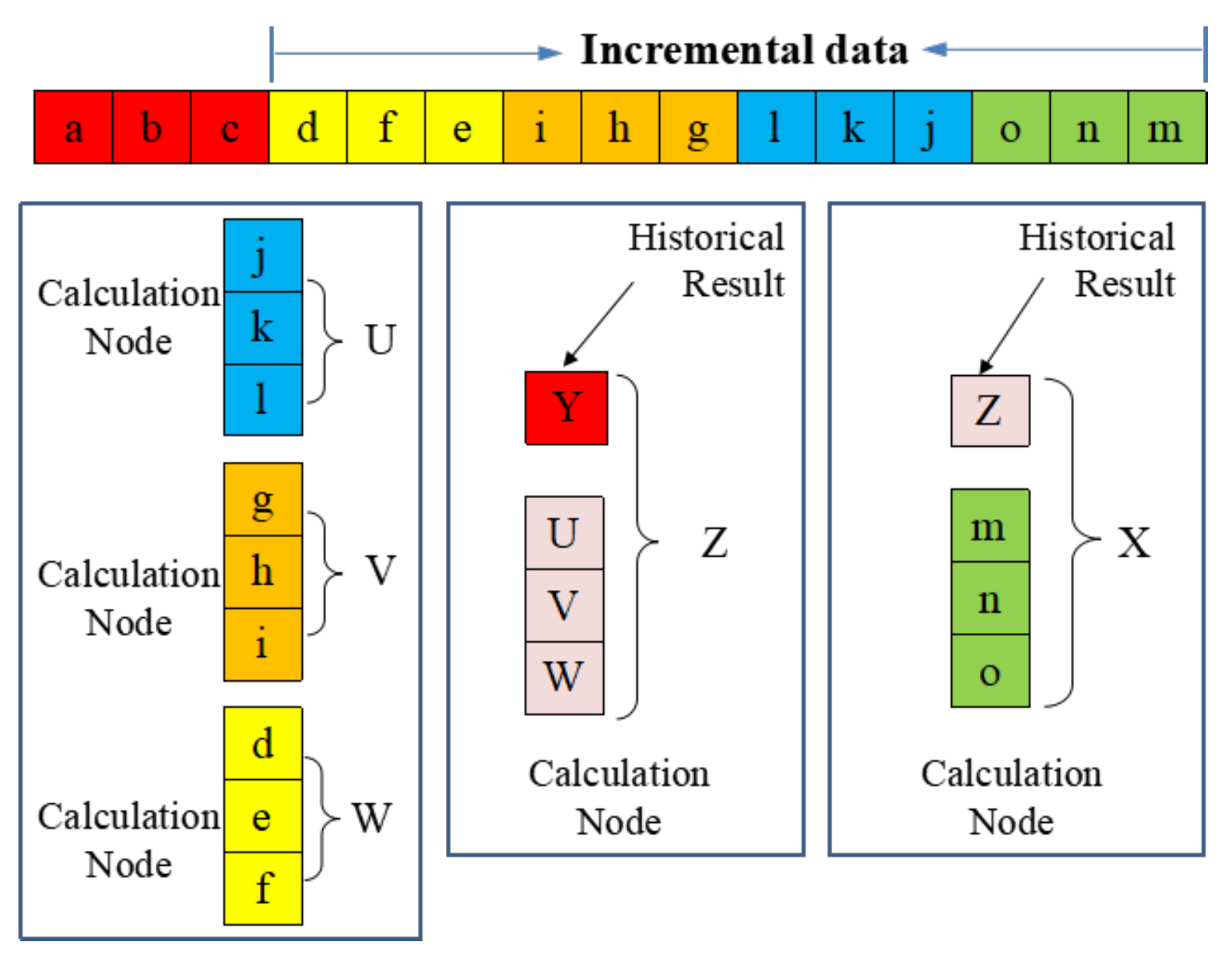}
        \caption{Challenge 3: decompose the calculation into the incremental computation}
        \label{fig:challenge3}
    \end{figure}

    The third challenge is the incremental computation of complex logic. It is of crucial importance to decompose the calculation of complex logic into the incremental scheme. In this way, it is possible to realize real-time computation. As in Figure \ref{fig:challenge3}, we own the results of the historical data. Thus we classify the incoming stream data as the incremental data. Then we eliminate the redundancy and group the stream data into different nodes. Each node stores its result as a partial result of the incremental computation. Up to now, we have implemented the real-time computation of several complex statistical metrics, such as variance, standard deviation, covariance, and K order central moment, etc.
    
    Last but not least, it is an excellent challenge of detecting the event sequences(complex events processing) based on the stream data. Usually, it is the easiest to recognize the user-defined event sequences inside the same dimension. It becomes harder when the event sequences occur across several dimensions. In the last, the event sequences based on the context are the most difficult.
    
    To overcome these challenges, the Stream Cube provides a fast dynamic data processing technology based on the incremental computation scheme. When the stream data comes, the Stream Cube decomposes the data along three axes: computation metric, data dimension, and time window respectively. Referring to Figure \ref{fig:streamcube}, we name the platform as the Stream Cube as each data unit is distributed along the three axes. The Stream Cube described above is an algorithm decomposing the complex logic of metrics into the incremental computation scheme. As each data unit is processed along the three axes, metrics can be calculated incrementally. Therefore, the Stream Cube becomes a distributed statistical computation platform, which supports multiple metrics such as counting, summing, average, max, min, variance, standard deviation, covariance, K order central moment as well as patterns of progressive increase and decrease, the discrimination of uniqueness and the detection of user-defined event sequences.

    \subsection{A Real-time intelligent data processing system}
    
    
    It has been several decades since AI techniques are applied in business scenarios for the first time. Generally, the applications can be divided into two classes: recommendation systems and fraud detection. For example, with the assistance of AI, CTR(click-through rate) of recommendation systems and FPR(false positive rate) of fraud detection can get significant improvement. However, restricted with the technology of stream data processing, there is still a considerable gap between real-time data processing and AI. Therefore, based on the Stream Cube, we deploy a system bringing AI into the real-time data processing.
    
    There are three general learning paradigms of AI(Artificial Intelligence). The first is formal methods, performing inference with the representation of symbolic logic. While it is highly interpretable, it requires expert knowledge. The second one, the statistical methods, also named as machine learning in computer science, has been applied to mine the intrinsic patterns of the data. Statistical methods are more flexible than formal methods, but it depends on the high quality of data. The last one, adopted from the cybernetics, is induced by problems and enhances itself from experience. 
    
    We have built various algorithms of three learning paradigms. The typical rule-based algorithms are the rule sets, the scorecard, and the knowledge graph. Also, there are several classical machine learning models, such as Random Forest, Decision Tree, Logistic Regression, and Deep Neural Network. So far, a real-time intelligent stream data processing system has been built upon the Stream Cube and AI.
    
    We can make a more precise judgment of business logic, rather than rules based on expert knowledge, by Artificial Intelligence. Classification problems are common in current business scenarios. For example, for better interpretability, rule sets are used to intercept suspicious behaviors in bank transactions. While for better performance, Gradient Boosting Descent Tree(GBDT) is usually preferred. With features computed by the Stream Cube, these algorithms can learn from variable data, and predict the risk of different behaviors. In this way, the model module can enable the system intelligence for various business scenarios.
    
\begin{figure}[!t]
        \centering
        \includegraphics[width=0.75\textwidth]{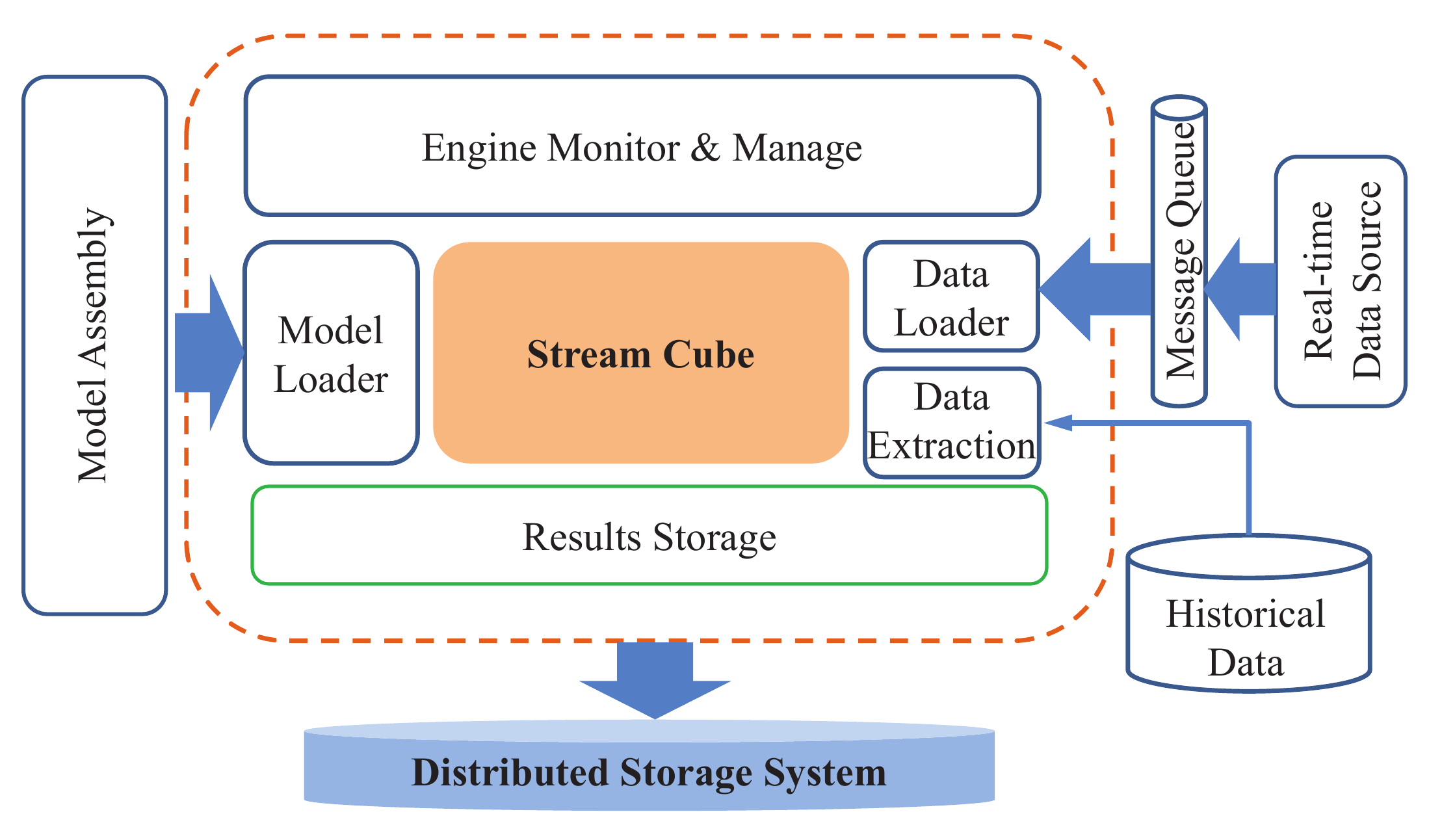}
        \caption{The architecture of the real-time intelligent data processing system}
        \label{fig:system}
    \end{figure}

Moreover, this kind of decision making can be realized as a service, called MaaS. We have proved the performance of the Stream Cube intelligent system in a few business scenarios. From the perspective of data flow, the system framework can be divided into three components in Figure\ref{fig:system}, the data source, the Stream Cube system and the data storage. In general, many business scenarios accumulate abundant production data, called historical data. The data source comes from both the historical data and the stream data of the production environment. The historical data will be processed by the data extraction module, while the real-time stream data will be fed into a real-time data message queue. These two ways provide the Stream Cube system with the data. Inside the system, the data loader handles the real-time data message queue and the data extraction module deals with the historical data. The core of the system engine consists of two modules, the computation engine, and the management system. The computation engine, namely the Stream Cube, is responsible for two kinds of computation requirements. One is the users' queries,  and the other is the input metrics of AI models. It is far faster and more powerful of the Stream Cube than existing streaming computation tools. The monitor and management system is a conventional module in system architecture. Firstly, it distributes various computation tasks and schedules the resources of the server cluster. Secondly, it monitors the whole system and handles different emergencies of the cluster. Besides, it is of great significance that the model assembly makes the system intelligent. In short words, we use PMML(Predictive Model Markup Language) to provide a unified model description interface for the system engine. Finally, all computation results will be stored in the distributed storage system.

    From the perspective of data flow, the system framework can be divided into three components as shown in Figure \ref{fig:system}: the data source, the intelligent stream data system, and the data storage. In practice, many business scenarios accumulate abundant production data, called historical data. The data source consists of both the historical data and the stream data in the production environment. Then, the historical data will be processed by the data extraction module, while the real-time data will be fed into a real-time data message queue. These two ways provide the Stream Cube system with the data. Above all, a monitor and management engine takes responsibility for the schedule of distributed tasks and recovery from failures. The intelligent system is a dual-core system. One core is the computation engine. The computation engine, namely the Stream Cube, is responsible for two kinds of computation requirements: the ad-hoc queries of analysts and the input metrics of AI models. The other core of the system is an AI module. Models are trained in offline mode with batching big data and used for real-time inference in the online production environment. In detail, we use PMML(Predictive Model Markup Language) to provide a unified model description interface for the intelligent system. Finally, all computation results will be stored in the distributed storage system.
    
    However, there are many practical problems in the implementation details. Firstly, in the industry, the origin data, either from the online transaction data or the data source of the third party, is usually disturbed by data noise. Therefore, a series of data cleaning, transformation, association, and integration tools have to be applied to enhance data quality. In the following, referring to the data flow, there are two branches for data processing, one is the real-time Stream Cube system, the other is batching big data processing platform. The real-time Stream Cube system computes predefined metrics, extracts the features, and builds the real-time hypergraph. The latter is integrated with the modeling and analysis platform to obtain the knowledge of data. From data acquisition to data processing, the data quality of the output metrics has been highly improved. With high-quality data, both business knowledge and users' preferences can be extracted to assist decision-making. The modeling and analysis platform is designed for modeling different scenarios and offers the algorithms of data mining, scoreboard generation, machine learning, and deep learning. The final decision maker can be a combination of multiple models, i.e., the set of rules, the scoreboard, the machine learning models and the knowledge graph.
    
    \section{Experiments}
    
    We have conducted some experiments to demonstrate the performance of our innovative computation engine, the Stream Cube. Firstly, the experiments evaluate the read and write performance of the Stream Cube system in the simultaneous computation of 16 metrics without processing business logic. Secondly, the experiments between the Stream Cube and Flink compare their performance under different circumstances.
    
    \subsection{Experiment Settings}
    
    \begin{table}
    \small
    \caption{Experiment Simulation Transaction Format}
    \label{tab:data-format}
    \begin{center}
        \begin{tabular}{ llp{10cm} }
            \hline\hline
            \textbf{Field }& \textbf{Type} & \textbf{Comment} \\ 
            \hline
            transTime & Long & Transaction Time \\ 
            acctId & String(32) & Transaction Initiator \\ 
            merId & String(32) & Transaction Receiver \\ 
            transAmt & Long & Transaction Limit \\ 
            city & String(32) & Transaction City \\ 
            hizCode & String(32) & Business Code. MOB-Telephone Recharging, 3C-3C Product, EXP-Expensive Goods, DIN-Dining, HOT-Hotel, OTH-Others \\ 
            chnl & String(3) & Transaction Channel. AND-Android Channel, IOS-Apple app Channel, WEB-PC Web Browser Channel, WAP-Mobile Phone Browser Channel \\ 
            stat & Integer & Transaction State. 0-Success, 1-Not Sufficient Funds \\ 
            \hline
        \end{tabular}
    \end{center}
    \end{table}
    
    Traditionally, the benchmark of the performance evaluation of the batching big data processing system is based on sorting. However, there has not been a publicly accepted performance benchmark of the stream data processing systems. Thus, we conducted a classical experiment of the performance of read and write when the number of servers varies from 1 to 8. In this experiment, the Stream Cube computes 16 statistical metrics simultaneously. The test environment consists of 8 servers, interconnected by 10G bandwidth. Each server has 12 core CPU and 256G memory.
    
    Then, we compare the performance between the Stream Cube and Flink by computing two statistical metrics respectively. The first metric is the average transaction amount of each user, with the changing time window size and a fixed window step of 1 day. The second metric is the number of unique transaction cities of each user, with the same settings of time window size and window step as the first metric. Software and their versions are listed below: JDK 1.8.0, Stream Cube 2.5.0, Aerospike 3.16.0.6, Flink 1.6.0, and Kafka 1.1.1. We test them on a single server of 8 core CPU and 128G memory.
    
    We simulate transactions in the production environment as in Table \ref{tab:data-format}. We generate the simulation data following rules below. The transaction time obeys Gaussian distribution and the transaction date varies in consecutive 150 days. There are 15 million records in total with 100,000 records each day. Both transaction initiators and receivers follow a Gaussian distribution of consumers. Then the transaction amounts conform to the Gaussian distribution ranging from CNY 0 to CNY 1,000,000. Lastly, we also generate transaction cities from the Gaussian distribution among 500 cites.
    
    \subsection{Results Analysis}
    
\begin{figure}[!t]
    \centering
    \scalebox{0.8}{%
        \begin{tikzpicture}
        \begin{axis}[
        xlabel={\textbf{Number of machines}},
        ylabel={\textbf{TPS (Times per Second)}},
        legend style={at={(0.4, 1)}, legend columns=1},
        bar width=.5cm,
        width=.7\textwidth,
        symbolic x coords={x1,x2,x4,x8},
        xtick=data,
        nodes near coords,
        ]
        \addplot table[x=machine, y=write]{\streamcube};
        \addplot table[x=machine, y=read]{\streamcube};
        \legend{Write, Read}
        \end{axis}
        \end{tikzpicture}
    }
    \caption{Write and read performance of the Stream Cube with up to 8 servers}
    \label{fig:stream-perf}
\end{figure}
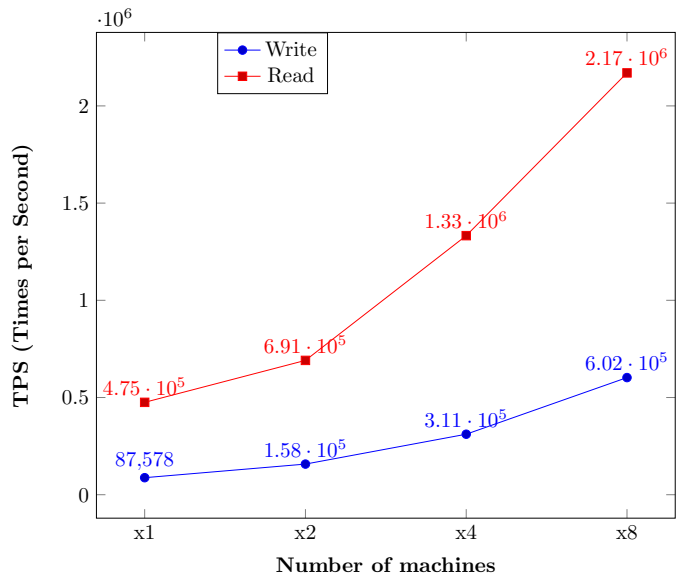

    In the first experiment, referring to the Figure  \ref{fig:stream-perf}, when running on a single node, the Stream Cube achieves more than 80,000 write TPS(times per second), and more than 400,000 read TPS. Besides, the average latency of these operations is between 1 and 2 milliseconds, which means the ability to realize the real-time computation of statistical metrics. It is well known that the write performance scales much slower than the read performance. However, due to its incremental computation scheme, both the write and read performance of the Stream Cube scales at a linear rate with numbers of servers increasing from 1 to 8. At last, the Stream Cube has not demonstrated its bottleneck with up to 8 servers, with over 2 million read TPS and over 600 thousand write TPS.
    
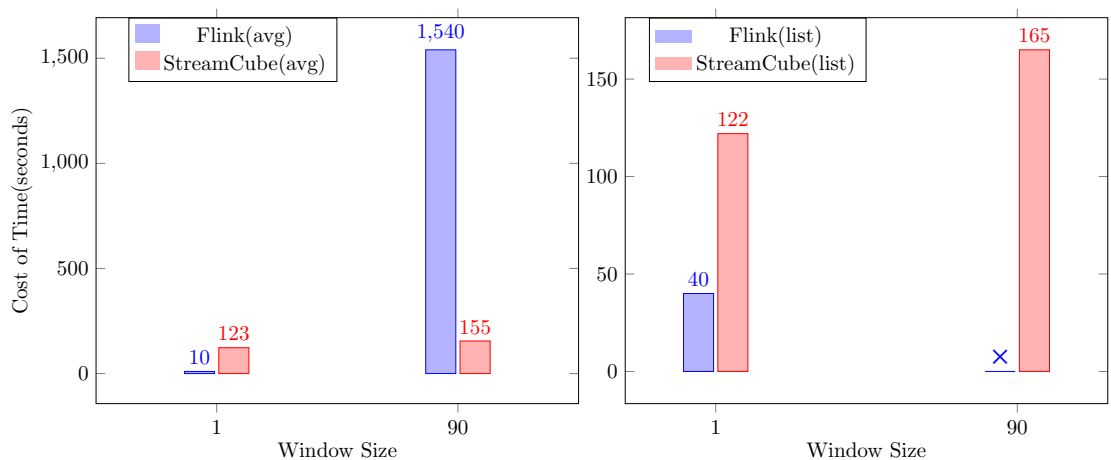
\begin{figure}[!t]
    \centering
    \scalebox{0.8}{%
        \begin{tikzpicture}
        \begin{axis}[
        name=plot1,
        ybar,
        width=6cm, height=8cm, enlarge x limits=0.5,
        xlabel={Window Size},
        ylabel={Cost of Time(seconds)},
        legend style={at={(0.5, 1)}, legend columns=1},
        legend image code/.code={%
            \draw[#1, draw=none] (0.2cm,-0.1cm) rectangle (0.6cm,0.2cm);
        },  
        bar width=.5cm,
        width=.6\textwidth,
        symbolic x coords={1,90},
        xtick=data,
        nodes near coords,
        ]
        \addplot table[x=window, y=flink-avg]{\compperf};
        \addplot table[x=window, y=stream-avg]{\compperf};
        \legend{Flink(avg), StreamCube(avg)}
        \end{axis}
        
        \begin{axis}[
        name=plot2,
        at=(plot1.right of south east),
        width=6cm, height=8cm, enlarge x limits=0.3,
        anchor=left of south west,
        ybar,
        xlabel={Window Size},
        legend style={at={(0.5, 1)}, legend columns=1},
        legend image code/.code={%
            \draw[#1, draw=none] (0cm,-0.1cm) rectangle (0.6cm,0.1cm);
        },  
        bar width=0.5cm,
        width=.6\textwidth,
        symbolic x coords={1,90},
        xtick=data, 
        nodes near coords*={%
     \pgfmathprintnumberto[fixed,assume math mode]{\pgfplotspointmeta}{\temp}%
     \ifthenelse{\temp=0}{\ding{53}}{\pgfmathprintnumber{\pgfplotspointmeta}}
   },
        ]
        \addplot table[x=window, y=flink-list]{\compperf};
        \addplot table[x=window, y=stream-list]{\compperf};
        \legend{Flink(list), StreamCube(list)}
        \end{axis}
        \end{tikzpicture}
    }
    \caption{Comparison of computational cost between Flink and Stream Cube under two different window sizes. Note that Flink(list) exhausted the memory when computing under 90 window size, we thus place a cross mark on the corresponding bar in the right figure.}
    \label{fig:comp-perf}
\end{figure}

    In the second experiment, as depicted in Figure  \ref{fig:comp-perf}, the initialization time of the Stream Cube is longer than Flink when the time window size is one day. Due to its pre-computed scheme, although the query window size is one day, the Stream Cube would divide all 150 days into disjoint computation intervals and then perform parallel computation. Therefore, with window size increasing to 90 days, given the first metric, the computation time of the Stream Cube increases very little, which is nearly one-tenth of that of Flink. As for the second metric, Flink spends more time when than that of the first metric the window size is one day, owing to the complexity of the second metric. However, the Stream Cube has no significant change because of its polynomial decomposition algorithm. However, when the window size is set to 90 days, Flink is unable to complete the computation of 90 days within 128 GB memory. Thus, we can conclude that the Stream Cube performs better than Flink in the second metric on a single node. 
    
    During the experiments, there are two drawbacks of Flink in practice. The one is the linear growth of computation time with the window size. The other is its inefficient hardware utilization: variable CPU utility, which is between 60\% and 90\%, and more memory requirement. In contrast, the Stream Cube shows computation efficiency with different window sizes and more stable CPU utility around 90\%. In a word, Flink is designed for stream data processing and loses its efficacy in the big data field. On the contrary, it is the incremental computation scheme and polynomial decomposition of statistical metrics that make the Stream Cube perform well at both the stream data processing and big data processing. 
    

\pgfplotstableread[row sep=\\,col sep=&]{
    machine & write     & read          \\
    x1        & 87578     & 475180     \\
    x2         & 157562     & 690986     \\
    x4        & 311232    & 1332560    \\
    x8        & 602498    & 2169778    \\
}\streamcube

\pgfplotstableread[row sep=\\,col sep=&]{
    window     & flink-avg & stream-avg & flink-list & stream-list         \\
    1        & 10     & 123 & 40 & 122     \\
    90         & 1540     & 155 & 0  & 165     \\
}\compperf

    \section{Applications}
    
    In this section, we show the capability of the real-time intelligent data processing system by two practical cases. The intelligent system focuses on the integration of real-time stream data processing and intelligent decision-making. The system targets the modeling analysis of various scenarios, which is towards practical applications. There are four critical abstract scenarios in our consideration, with the help of the low latency of data processing, the capability of a vast amount of data, and AI(Artificial Intelligence) models of the system. The first scenario is the fingerprint of devices, aiming at discriminating real devices from the fake ones. The second one is the machine defense, the goal of which is resisting instantaneous peak visit flows. Then, the data service, the third case, provides enterprises with high-quality data. Last but not least, the ability to tell computers and humans apart plays an essential role in current Internet business models like online advertising, and live streaming, etc.
    
    \begin{table}
    \small
    \caption{The stream data volume in various applications}
    \label{tab:data-volume}
    \begin{center}
        \begin{tabular}{lll}
            \hline\hline
            \textbf{Field }& \textbf{Scenario} & \textbf{Stream Data Volume(per second)} \\ 
            \hline
            Finance & China UMS & 50,000 transactions \\
            E-Commerce & Taobao 11.11 Day & 220,000 transactions \\
            IoT & Shanghai Railway Line 1 & 200,000 records \\
            Ticket & 12306 Railway Ticket & 1.7 million page views \\
            \hline
        \end{tabular}
    \end{center}
    \end{table}
    
    The above abstract scenarios are motivated by the increasing need for data processing technology, the real-time analysis methods, and the intelligent decision-making system. With the continuous development of the economy of China, cable broadband and the mobile Internet have been particularly popular among people. Therefore, the e-commerce industry has significant progress. More and more fields have been electronic from buying railway tickets, online shopping to take-out service and bicycle sharing. Besides, public facilities like the underground have to monitor the real-time running status of the system. While the growing industry provides convenience for citizens, it is hard to maintain such a large-scale system, serving for millions or even billions of people. 
    
    Based on the four abstract scenarios, there are broad practical business applications. Firstly, the military industry and Internet security both require specific solutions to big data processing. Then, the IoT(Internet of Things) such as the railway transmit system has to deal with vast amounts of data each second. Lastly, many industry fields need the enhancement of artificial intelligence, like intelligent manufacturing, financial risk management, the smart city, and public security, etc.
    
    As shown in Table \ref{tab:data-volume}, there are full of challenges nowadays. The China UMS, the largest payment institution of China, processes 50,000 transactions of bankcard payment per second. Moreover, in double 11(Singles' Day), the most significant online shopping day in China, the peak trading volume of Taobao mall can reach 220,000 transactions per second. Besides the finance and e-commerce, the stream data volume cannot be underestimated in the IoT(Internet of Things) and the ticket business. The only 200 sensors of the Metro Line 1 in Shanghai produces 200,000 records per second. 

Also, 12306 is the national unified railway ticket institution. Consequently, the page views of the 12306 websites can even reach 1,700,000 times per second in the annual ticket fight during the Spring Festival travel rush. So far, it has already been a hard problem to maintain large-scale business service. Furthermore, the task to realize a real-time intelligent data processing technology system of the stream big data seems even impossible.
    
    Current systems can only merely support such vast amounts of business transactions. It is of extreme difficulty but also of the great need to realize real-time risk management, anomaly detection, and intelligent decision-making. For example, in financial transactions, it saves massive economic losses to detect bankcard fraud in time. In e-commerce, system vulnerabilities can be used to damage the profits of enterprises. In IoT(Internet of Things), human's security can be in danger if hackers try to break into private facilities. Therefore, we provide two practical examples where the intelligent data processing system is applied in the production environment.
    
    The first application is bankcard fraud detection in risk management of the finance field. Before the equipment of the Stream Cube system for risk management, there are three channels of transactions, the electronic payment, the direct bank, and Internet finance. Then the ESB system processes vast amounts of transactions and stores them in the enterprise big data system, which is replicated into the Hadoop and HBase in real time.
    
    To obtain the real-time data, synchronous and asynchronous probes are both placed in front of the commerce system. Besides, third-party cloud data is also introduced to enhance the performance of the risk management system. As for the knowledge of risk management, hundreds of rules, coming from the machine learning models and the expert experience, have been distilled for risk management engine. After the initial detection of these rules, transactions with high risk are fed to the stream data engine to compute current risk metrics. The results of high-risk transactions are then reversely supplied with the risk management engine for further analysis. In the production environment, with only four servers, risk management engine responses in less than 50 milliseconds and processes over 50,000 transactions per second. Risk management backend will verdict the approval or rejection of the risk transactions according to the results of the risk management engine.
    
    The second case is the application of detecting the abnormal behaviors of buying railway tickets in 12306. In China, 12306 is responsible for selling train tickets to billions of people. Thus, its peak page views can reach the order of hundreds of billions in one day. To prevent the ticket scalpers from disturbing the regular market order, a new image recognition CAPTCHA has been invented. According to the given keywords, people have to recognize the right images from the image list. However, the ticket scalpers are far powerful than ordinary passengers so that they tackle the problematic CAPTCHA with crowdsourcing. On the contrary, regular passengers pass the CAPTCHA test much slower than the ticket scalpers. 
    
    A detailed analysis of the visiting traffic demonstrates that, in 12306, real users contribute 30 million visits per day in peacetime and spiders developed by programming experts account for 1.5 billion visits per day, 50 times of that of real users. At the peak moment, real users visit 30 billion times per day, while spiders visit 150 billion times per day, about 1.7 million times per second. It is far more difficult to discriminate the ticket scalpers precisely in 0.1 seconds with modeling the past behaviors for every buyer than the fraud detection in risk management of Internet finance.
    
    \section{Conclusion}
    
    Batching big data has been mined a lot to improve business performance. The knowledge of business logic can be updated based on batching big data mining technologies. However, the real-time data is the most valuable to either reflect the change of the preference of consumers or describe the pattern shift of fraudsters. It is more meaningful to promote real-time stream data technology.
    
    In this paper, we propose an innovative solution of the stream data processing, the Stream Cube, based on the incremental scheme of calculation and polynomial decomposition of metrics. To achieve the goal of intelligent decision-making, we have also built an intelligent data processing system supporting the Stream Cube for real-time processing and AI models for data mining. Then, the two applications have proved the performance of the intelligent system.
    
    However, there is no endpoint pursuing the improvement of business performance. On the one hand, the accurate recognition of abnormal behaviors of 12306 remains a problem. On the other hand, the computation and inference of the next generation of real-time hypergraph have not been solved yet. There are still full of challenges and opportunities in the future.
    
    \Acknowledgements{}

    

 

\begin{thebibliography}{99}
    	
    	\bibitem{1} Pan Y. Heading toward artificial intelligence 2.0. Engineering, 2016, 2: 409-413
		\bibitem{2} Chen C. Real-time processing technology, platform and application of streaming big data. Big Data, 2017, 3: 1-8
		\bibitem{3} Shvachko K, Kuang H, Radia S, et al. The hadoop distributed file system. In Mass storage systems and technologies (MSST), 2010. 1-10
		\bibitem{4} Dean J, Ghemawat S. MapReduce: simplified data processing on large clusters. Communications of the ACM, 2008, 51: 107-113
		\bibitem{5} Zaharia M, Chowdhury M, Franklin M J, et al. Spark: Cluster computing with working sets. HotCloud, 2010, 10: 95
		\bibitem{6} Zhang Q, Cheng L, Boutaba R. Cloud computing: state-of-the-art and research challenges. Journal of internet services and applications, 2010, 1: 7-18
		\bibitem{7} Hashem I A, Yaqoob I, Anuar N B, et al. The rise of “big data” on cloud computing: Review and open research issues. Information systems, 2015, 47: 98-115
		\bibitem{8} Wu Q, Ishikawa F, Zhu Q, et al. Deadline-constrained cost optimization approaches for workflow scheduling in clouds. IEEE Transactions on Parallel and Distributed Systems, 2017, 28: 3401-3412
		\bibitem{9} Saha B, Shah H, Seth S, et al. Apache tez: A unifying framework for modeling and building data processing applications. In Proceedings of the 2015 ACM SIGMOD international conference on Management of Data, 2015. 1357-1369
		\bibitem{10} Maarala A I, Rautiainen M, Salmi M, et al. Low latency analytics for streaming traffic data with Apache Spark. In Big Data (Big Data), 2015 IEEE International Conference on,  2015. 2855-2858
		\bibitem{11} Toshniwal A, Taneja S, Shukla A, et al. Storm@ twitter. In Proceedings of the 2014 ACM SIGMOD international conference on Management of data, 2014. 147-156
		\bibitem{12} Carbone P, Katsifodimos A, Ewen S, et al. Apache flink: Stream and batch processing in a single engine. Bulletin of the IEEE Computer Society Technical Committee on Data Engineering, 2015, 36: 4
		\bibitem{13} Zaharia M, Das T, Li H, et al. Discretized Streams: An Efficient and Fault-Tolerant Model for Stream Processing on Large Clusters. HotCloud, 2012, 12: 10
		\bibitem{14} Zhao X, Garg S, Queiroz C, Buyya R. A Taxonomy and Survey of Stream Processing Systems. In Software Architecture for Big Data and the Cloud, 2017. 183-206
		\bibitem{15} Ali M. An introduction to microsoft sql server streaminsight. In Proceedings of the 1st International Conference and Exhibition on Computing for Geospatial Research \& Application , 2010. 66
		\bibitem{16} Hyde J. Data in flight. Communications of the ACM, 2010, 53: 48-52
		\bibitem{17} StreamBase I. Streambase: Real-time, low latency data processing with a stream processing engine
		\bibitem{18} Demers A J, Gehrke J, Panda B, et al, White WM. Cayuga: A General Purpose Event Monitoring System. In Cidr, 2007, 7: 412-422
		\bibitem{19} Strohbach M, Ziekow H, Gazis V, et al. Towards a big data analytics framework for IoT and smart city applications. In Modeling and processing for next-generation big-data technologies, 2015. 257-282
		\bibitem{20} Noghabi S A, Paramasivam K, Pan Y, et al. Samza: stateful scalable stream processing at LinkedIn. Proceedings of the VLDB Endowment, 2017, 10: 1634-45
		\bibitem{21} Chauhan J, Chowdhury S A, Makaroff D. Performance evaluation of Yahoo! S4: A first look. In 2012 Seventh International Conference on P2P, Parallel, Grid, Cloud and Internet Computing, 2012. 58-65
		\bibitem{22} Fernandez R C, Pietzuch P R, Kreps J, et al. Liquid: Unifying Nearline and Offline Big Data Integration. In CIDR, 2015
		\bibitem{23} Pacaci A, Özsu M T. Distribution-Aware Stream Partitioning for Distributed Stream Processing Systems. In Proceedings of the 5th ACM SIGMOD Workshop on Algorithms and Systems for MapReduce and Beyond, 2018. 6
		\bibitem{24} Jin H, Chen F, Wu S, et al. Towards Low-Latency Batched Stream Processing by Pre-Scheduling. IEEE Transactions on Parallel and Distributed Systems, 2018
		\bibitem{25} Venkataraman S, Panda A, Ousterhout K, et al. Drizzle: Fast and adaptable stream processing at scale. In Proceedings of the 26th Symposium on Operating Systems Principles, 2017. 374-389
		\bibitem{26} Zhang B, Jin X, Ratnasamy S, et al. Awstream: Adaptive wide-area streaming analytics. In Proceedings of the 2018 Conference of the ACM Special Interest Group on Data Communication, 2018. 236-252
		\bibitem{27} Li W, Niu D, Liu Y, et al. Wide-area spark streaming: Automated routing and batch sizing. IEEE Transactions on Parallel and Distributed Systems, 2018
		\bibitem{28} Traub J, Grulich P M, Cuellar A R, et al. Scotty: Efficient Window Aggregation for out-of-order Stream Processing. In2018 IEEE 34th International Conference on Data Engineering, 2018. 1300-1303
		\bibitem{29} Srinivasan V, Bulkowski B, Chu W L, Set al. Aerospike: architecture of a real-time operational DBMS. Proceedings of the VLDB Endowment, 2016, 9: 1389-400
		\bibitem{30} Carlson J L. Redis in action. Manning Publications Co, 2013
		
    \end{thebibliography}
    
    
    
\end{document}